\documentclass[a4paper, amsfonts, amssymb, amsmath, reprint, showkeys, nofootinbib, twoside,notitlepage,onecolumn]{revtex4-1}
\def\apj{{ApJ}}                 % Astrophysical Journal
\def\mnras{{MNRAS}}             % Monthly Notices of the RAS
\def\prd{{Phys.~Rev.~D}}        % Physical Review D
\usepackage{amsmath,amstext}
\usepackage[T1]{fontenc}
\usepackage{amssymb}
\usepackage{graphicx}
\usepackage{ae,aecompl}

\DeclareFontFamily{OT1}{pzc}{}
\DeclareFontShape{OT1}{pzc}{m}{it}{<-> s * [1.10] pzcmi7t}{}
\DeclareMathAlphabet{\mathpzc}{OT1}{pzc}{m}{it}

\usepackage{hyperref}
\usepackage{amsmath}
\usepackage{amssymb}
\usepackage{mathtools}
\usepackage{bm}
\usepackage{cleveref}
\usepackage{tensor}
\usepackage{braket}
\usepackage{enumitem}
\usepackage{mhchem}
\usepackage{amsthm}
\usepackage{nccmath}
\usepackage{mathrsfs}
\usepackage{color}

\newcommand{\spc}{\quad \quad \quad}

\def\be{\begin{equation}}
\def\ee{\end{equation}}
\def\beq{\begin{eqnarray}}
\def\eeq{\end{eqnarray}}

\theoremstyle{definition}

\theoremstyle{theorem}

\begin{document}
\title{Stationary Couette-type flows in relativistic fluids}
\author{L.~Gavassino$^1$, P.~Niekamp$^2$, S.~Schlichting$^3$, and G.S.~Denicol$^4$}
\affiliation{$^1$Department of Applied Mathematics and Theoretical Physics, University of Cambridge, Wilberforce Road, Cambridge CB3 0WA, United Kingdom\\
$^2$Institut für Theoretische Physik, Justus-Liebig-Universit\"{a}t, 35392 Giessen, Germany\\
$^3$Fakult\"{a}t f\"{u}r Physik, Universit\"{a}t Bielefeld, D-33615 Bielefeld, Germany\\
$^4$Instituto de Física, Universidade Federal Fluminense Av. Gal. Milton Tavares de Souza, S/N, 24210-346, Gragoatá, Niterói, Rio de Janeiro, Brazil}

\begin{abstract}
We investigate a class of stationary, planar-symmetric solutions of relativistic hydrodynamics, in which a dissipative fluid is confined between two parallel plates that move relative to each other and/or are maintained at different temperatures. We find that neglecting the heat flux leads to qualitatively incorrect flow profiles, even in systems with temperature-independent viscosity. This arises from the fact that, in special relativity, the heat flux itself contributes to the momentum density (the so-called ``inertia of heat''). This effect is most evident in the Landau frame, where the fluid removes the excess energy generated by viscous heating by streaming across the boundaries. The analysis is further extended to the limit of vanishing chemical potential. 
\end{abstract} 
\maketitle

\section{Introduction}

\begin{figure}[b!]
    \centering
\includegraphics[width=0.45\linewidth]{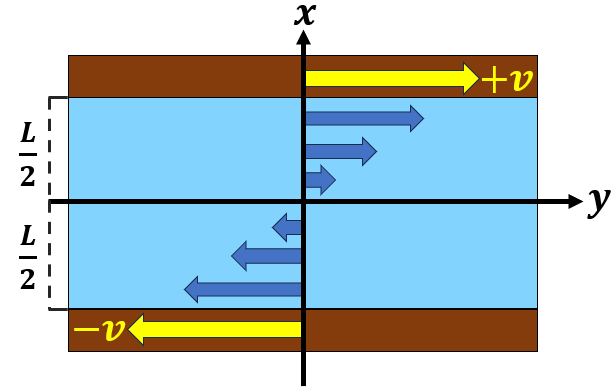}
\includegraphics[width=0.45\linewidth]{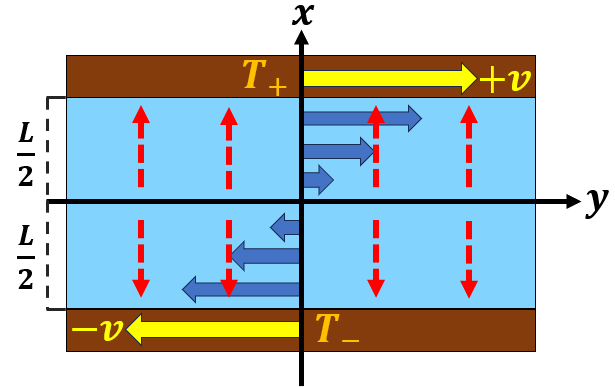}
    \caption{Left panel: Classical Couette flow. Two parallel plates (brown) are in relative motion, forcing a fluid confined between them (blue) to undergo a shear flow. Right panel: Generalization that accounts for heat exchanges. Now the plates have some temperatures $T_\pm$ (often equal), and they can absorb the excess heat (red arrows) generated by viscous dissipation.}
    \label{fig:sketch}
\end{figure}

One of the best-known stationary solutions of the Navier-Stokes equations is the Couette flow \cite{SchlichtingBoundaryLayerBook2017},
\begin{equation}\label{couetteNewtonian}
(u^x, u^y, u^z) = (0, 2vx/L, 0)\,,
\end{equation}
which describes the velocity field of a viscous fluid of uniform viscosity, confined between two infinite parallel plates located at $x = \pm L/2$ and moving in opposite directions with velocities $v_y = \pm v$ (see Fig.~\ref{fig:sketch}, left panel). Given the universality and pedagogical importance of this flow, it is natural to ask what new features arise once relativistic effects are included. Intuitively, one expects deviations from Eq.~\eqref{couetteNewtonian} to emerge as $v$ approaches unity, through Lorentz factors and alike. This expectation was indeed confirmed by \citet{Rogava1996}, who derived a relativistic version of \eqref{couetteNewtonian} by solving the conservation law of the $y$ component of the momentum (i.e.
$\partial_\mu T^{\mu y} = 0$, with $T^{\mu \nu}$ the stress-energy tensor)
for a stationary and planar-symmetric fluid with constant viscosity and vanishing heat conductivity, obtaining
\begin{equation}\label{couetteVisco}
(u^0, u^x, u^y, u^z)
= \left(
\sqrt{1 + \frac{(2vx/L)^2}{1 - v^2}},
0,
\frac{2vx/L}{\sqrt{1 - v^2}},
0
\right) .
\end{equation}
Unfortunately, Rogava's decision to set the heat conductivity to zero has the unwanted consequence that the energy conservation law ($\partial_\mu T^{\mu t}{=}\,0$) is not satisfied exactly. In fact, fluids undergoing shear necessarily warm up by viscous heating, and steady states can be maintained only if the excess energy is expelled through the boundaries, which requires a finite thermal conductivity. Once the conductivity is restored, however, new relativistic corrections arise, because the heat flux carries momentum \cite[\S 5.7]{MTW_book}, so it enters directly $\partial_\mu T^{\mu y}{=}\,0$, thereby modifying \eqref{couetteVisco}.

\newpage

The purpose of this work is to investigate the corrections to Eq.~\eqref{couetteVisco} that arise when thermal conduction is taken into account. To this end, we generalize the classical Couette setup by imposing fixed temperatures $T_{\pm}$ at the plates, and allowing heat flux to pass through the fluid-boundary interface (see Fig.~\ref{fig:sketch}, right panel). A first analysis of this type was presented in Ref.~\cite{Granik1996}, where these generalized Couette profiles were calculated in the case of a fluid with finite chemical potential modeled within Eckart’s first-order framework~\cite{Eckart40}. In the present study, we expand that treatment by (a) discussing in detail the role of the ``inertia of heat'', (b) examining the differences in behavior between the Eckart and the Landau frame \cite{landau6,OlsonLifsh1990}, and (c) analyzing the limit of vanishing chemical potential.

Throughout the article, we adopt the metric signature $(-,+,+,+)$, and work in natural units, $c=\hbar=k_B=1$.

\vspace{-0.2cm}
\section{Generalized Couette problem at finite chemical potential}\label{finitUdine}
\vspace{-0.2cm}

In this section, we solve the generalized Couette problem (Fig.~\ref{fig:sketch}, right panel) for a fluid with finite chemical potential. That is, for a fluid that carries a conserved particle current $J^\mu$. As we shall see, the existence of this additional current renders the analysis considerably simpler, provided that the particles cannot cross the boundaries.

\vspace{-0.2cm}
\subsection{Convenience of the Eckart frame}\label{Eckartiamo!}
\vspace{-0.2cm}

Under the assumptions of stationarity ($\partial_t = 0$) and planar symmetry ($\partial_y = \partial_z = 0$), the particle conservation law simplifies to  $0 = \partial_\mu J^\mu = \partial_x J^x$,  
implying that $J^x$ is constant throughout the flow. Imposing the boundary condition that the particle current be tangential to the plates (i.e. no particles cross the boundaries), we obtain
\begin{equation}\label{Jxxxx}
J^x = 0 \, . 
\end{equation}  
It follows that, if we work in the Eckart frame ($u^\mu \propto J^\mu$), the four-velocity takes the form
\begin{equation}
u^\mu =
\begin{bmatrix}
\gamma(x)\\
0\\
u(x)\\
0 \\
\end{bmatrix} \, ,
\end{equation}
with $\gamma^2-u^2=1$. This result is very convenient, because it tells that contracting a velocity index with a derivative index always returns zero, in particular
\begin{equation}\label{dxuxiszero}
u^\mu \partial_\mu =u^x \partial_x =0 \, , \spc \partial_\mu u^\mu =\partial_x u^x=0  \, .
\end{equation}
Notably, this would no longer be true in the Landau frame \cite{landau6}. Because heat can flow across the boundaries, and the Landau velocity follows the energy transport, one generally finds $u_L^x \neq 0$. For this reason, it is most convenient to carry out the calculation entirely in the Eckart frame, and then transform the final solution to the Landau frame. This will be our strategy.

\vspace{-0.2cm}
\subsection{Energy-momentum conservation}
\vspace{-0.2cm}

In the Eckart frame, the stress-energy tensor can be geometrically decomposed as follows:
\begin{equation}\label{TimunuF}
T^{\mu \nu}=\varepsilon u^\mu u^\nu +(P+\Pi)\Delta^{\mu \nu}+u^\mu q^\nu +u^\nu q^\mu +\pi^{\mu \nu} \, ,
\end{equation}
with $\varepsilon$ the energy density, $P$ the equilibrium pressure, $\Delta^{\mu \nu}=\eta^{\mu \nu}+u^\mu u^\nu$ the projector orthogonal to the velocity ($\Delta^{\mu\nu}u_\nu=0$), $\Pi$ the bulk-viscous pressure, $q^\mu$ the heat flux (satisfying $q^\mu u_\mu =0$), and $\pi^{\mu \nu}$ the shear stress tensor (satisfying $\pi^{\mu \nu}u_\nu =0$ and $\pi^\mu_\mu=0$).  Then, local energy-momentum conservation gives
\begin{equation}\label{ABC}
\begin{split}
\partial_x T^{xt}=0 & \qquad \Longrightarrow \qquad T^{xt}=\gamma q^x +\pi^{xt}=A \, , \\
\partial_x T^{xx}=0 & \qquad \Longrightarrow \qquad T^{xx}=P+\Pi  +\pi^{xx}=B \, , \\
\partial_x T^{xy}=0 & \qquad \Longrightarrow \qquad  T^{xy}=u q^x +\pi^{xy}=C \, .\\
\end{split}
\end{equation}
By looking at this system, we immediately learn two things. First, we see that, indeed, there is a relativistic correction $\partial_{x}(uq^x)$ to the conservation law $\partial_x T^{xy}=0$, confirming that the heat flux modifies the shape of the flow by contributing to the fluid's inertia (i.e. to its momentum flux). Secondly, we also see that we cannot set $q^x=0$. In fact, if we do it, we obtain that both $\pi^{xt}$ and $\pi^{xy}$ are constant. But these components of the stress tensor are related by the geometrical constraint $0=\pi^{x\nu}u_\nu=-\pi^{xt}\gamma+\pi^{xy}u$, which means that the only possible solution would be $u=\text{const}$, i.e. no shear.
\newpage
Using the constraint $\pi^{xt}\gamma=\pi^{xy}u$, we can isolate $q^x$, $P$, and $\pi^{xy}$, giving
\vspace{-0.1cm}
\begin{equation}\label{prima}
\begin{split}
q^x={}& \gamma A- uC \, ,\\
P={}& B-\Pi -\pi^{xx} \, , \\
\pi^{xy}={}& \gamma^2 C-\gamma u A \, .\\
\end{split}
\end{equation}

\vspace{-0.3cm}
\subsection{Israel-Stewart relaxation equations}
\vspace{-0.3cm}

Since the relativistic Navier-Stokes theory of Eckart \cite{Eckart40} is acausal and unstable \cite{Hiscock_Insatibility_first_order}, the corresponding initial-value problem is ill-posed \cite{Kost2000,GavassinoSuperlum2021}, and physical solutions cannot be distinguished \emph{a priori} from spurious ones. We therefore introduce Israel-Stewart-type relaxation equations for the dissipative fluxes \cite{Israel_Stewart_1979} as a technical regulation scheme:
\begin{equation}\label{Israeliamo!}
\begin{split}
\Pi={}& -\zeta \partial_\mu u^\mu-\tau_\Pi u^\mu \partial_\mu \Pi \, , \\
q^\nu={}& -\Delta^{\nu}_\alpha[\chi( \partial^\alpha T +T u^\mu \partial_\mu u^\alpha)+\tau_q u^\mu \partial_\mu q^\alpha] \, , \\
\pi^{\nu\rho}={}& -\Delta^{\nu \rho}_{\alpha \beta}[2\eta \partial^\alpha u^\beta +\tau_\pi u^\mu \partial_\mu \pi^{\alpha\beta}] \, . \\
\end{split}
\end{equation}
where $\{\zeta,\chi,\eta\}$ are the bulk viscosity, heat conductivity, and shear viscosity, $\{\tau_\Pi,\tau_q,\tau_\pi\}$ are relaxation times, and $\Delta^{\nu \rho}_{\alpha \beta}=\frac{1}{2}(\Delta^\nu_\alpha \Delta^\rho_\beta +\Delta^\nu_\beta \Delta^\rho_\alpha-\frac{2}{3}\Delta^{\nu \rho}\Delta_{\alpha\beta})$ extracts the symmetric, transverse, traceless part of a two-tensor. Thanks to Eq. \eqref{dxuxiszero}, we find that all terms involving relaxation times vanish, so we effectively recover Eckart's theory. Hence, the regulator terms we introduce to restore well-posedness do not affect the profile, which is therefore robust. Then, we find that, in the Couette geometry, $\Pi\,{=}\,\pi^{xx}{=}\,0$, $\pi^{xy}{=}\,{-}\eta\partial_x u$, and $q^x{=}{-}\chi \partial_x T$. Thus, the system \eqref{prima} becomes
\begin{equation}\label{dopo}
\begin{split}
\chi \partial_x T={}& uC-\gamma A \, ,\\
P={}& B \, , \\
\eta \partial_x u={}& \gamma u A-\gamma^2 C \, .\\
\end{split}
\end{equation}
such that the pressure $P$ remains at a constant value $B$ throughout the fluid. Finally, if we further assume that  the viscosity $\eta(P,T)$ is independent of $T$ (see App. \ref{AppAAA} for an example where $\eta$ depends on $T$), then we can define some new constants $\Bar{A}\equiv \frac{A}{\eta(P=B)}$ and $\Bar{C}\equiv \frac{C}{\eta(P=B)}$, and the third line of \eqref{dopo} reduces to 
\begin{equation}\label{thepowerful}
\partial_x u =\gamma u \Bar{A} -\gamma^2 \Bar{C}\, .
\end{equation}
We have uncovered a rather surprising result. Although a nonzero thermal conductivity is required for internal consistency, and despite the fact that the heat flux explicitly appears in the momentum conservation law, the resulting velocity profile $u(x)$ does not depend on the exact value of the conductivity (except via a temperature-dependent $\eta$). In this sense, the heat-related corrections to Eq.~\eqref{couetteVisco} exhibit a \textit{universal form}.

\vspace{-0.2cm}
\subsection{Universal symmetric solution}\label{Thelasymmetry}
\vspace{-0.1cm}

\begin{figure}[h!]
    \centering
\includegraphics[width=0.49\linewidth]{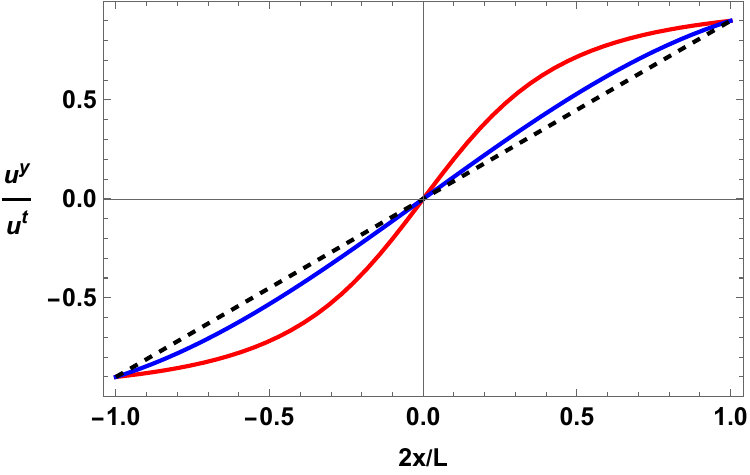}
\includegraphics[width=0.49\linewidth]{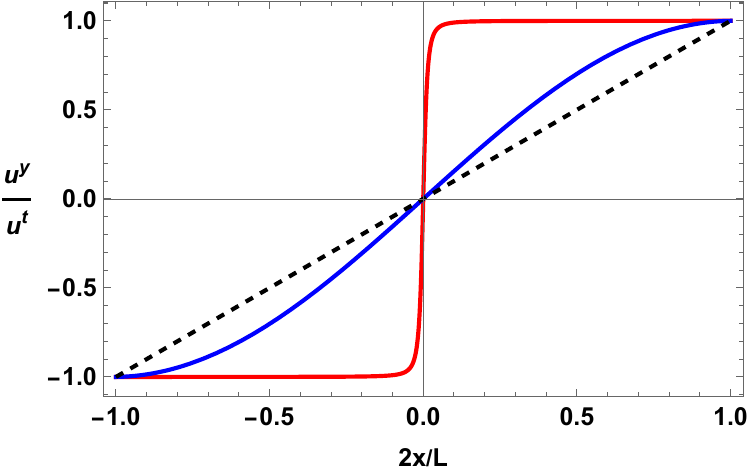}
\caption{Velocity profile of a relativistic fluid with uniform viscosity undergoing a generalized Couette flow (see Fig.~\ref{fig:sketch}), for equal boundary temperatures ($T_+=T_-$) and with plate velocities $v=0.9$ (left panel) and $v=0.9999$ (right panel). The solid blue line shows the exact solution~\eqref{exactone}, the red line Rogava's model \eqref{rogG} (which neglects the heat flux \cite{Rogava1996}), and the dashed line the Newtonian result, where the speed increases linearly with $x$. }
\label{fig:speeddoni}
\end{figure}

Let us specialize Eq. \eqref{thepowerful} to the case where $T_+=T_-$. Then, the problem is symmetric under $180^{\text{o}}$-rotations around the $z$ axis, i.e. under the transformation $(x,y,z)\rightarrow (-x,-y,z)$. If we apply this symmetry to \eqref{ABC}, we find that $A=0$. Therefore, \eqref{thepowerful} reduces to 
\vspace{-0.2cm}
\begin{equation}
\partial_x u = -(1+u^2) \Bar{C}\, .
\end{equation}
This is a separable ordinary differential equation, which can be solved analytically. By imposing the boundary conditions $u(\pm L/2)=\pm v/\sqrt{1{-}v^2}$, we can determine $\bar{C}$ and finally obtain
\begin{equation}\label{exactone}
u(x)=\tan\left[\dfrac{2x}{L} \arctan\left(\dfrac{v}{\sqrt{1{-}v^2}} \right) \right] \, .
\end{equation}
By comparison, \citet{Rogava1996}, who neglected the effect of heat conduction, ended up solving the simplified equation $\partial_x u = -\Bar{C}$, and only obtained
\vspace{-0.1cm}
\begin{equation}\label{rogG}
u(x)=\dfrac{2vx}{L\sqrt{1 {-} v^2}} \qquad (\text{Rogava's approximation})\, .
\end{equation}
In Fig.~\ref{fig:speeddoni}, we compare the predictions of Eqs.~\eqref{exactone} and \eqref{rogG}. As shown, neglecting the heat flux leads to a substantial overestimation of the flow velocity, rendering Rogava’s approximation even less accurate than the Newtonian result. This discrepancy is particularly striking in the limit $v \,{\to}\, 1$, where Rogava’s  approximate solution diverges everywhere (except at $x\,{=}\,0$), while the exact solution remains finite throughout the domain (except at the boundary), approaching
\begin{equation}
u(x) \xrightarrow{v \to 1} \tan\left( \frac{\pi x}{L} \right) .
\end{equation}

\subsection{Temperature profile and heat flux}

\begin{figure}[b!]
    \centering
\includegraphics[width=0.49\linewidth]{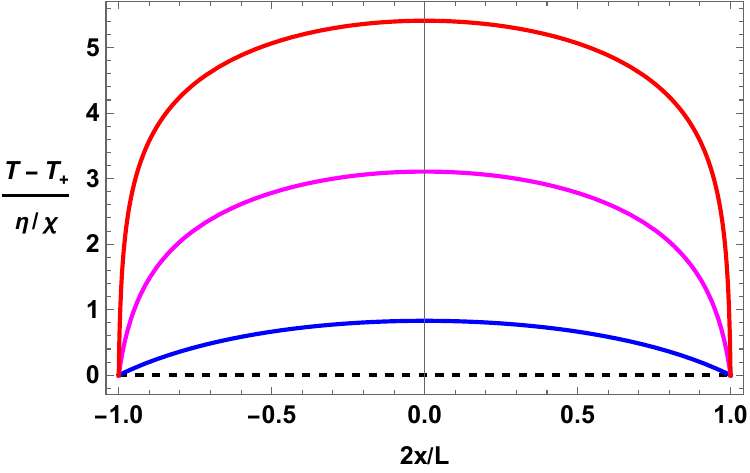}
\includegraphics[width=0.49\linewidth]{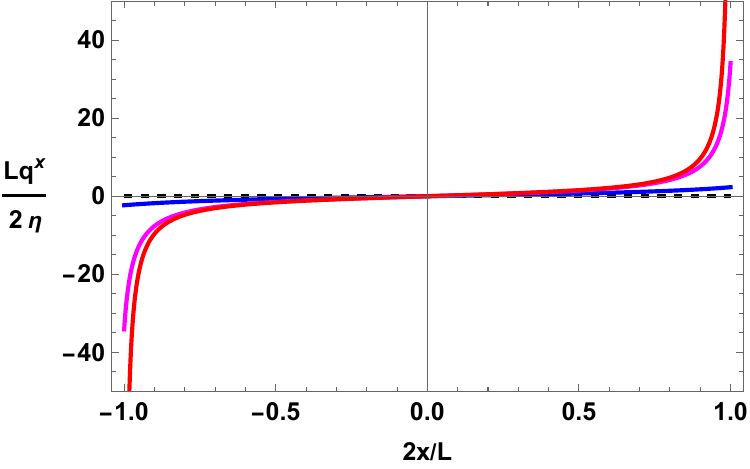}
\caption{Temperature (left panel) and heat flux (right panel) profiles of a relativistic fluid with uniform $\eta$ and $\chi$ undergoing a generalized Couette flow (see Fig.\ref{fig:sketch}), for equal boundary temperatures ($T_+=T_-$), and with plate velocities $v=0$ (dashed), $0.9$ (blue), $0.999$ (magenta), and $0.99999$ (red). The analytical expressions are provided in Eq. \eqref{temperatueandheattone}.}
\label{fig:temperone}
\end{figure}

Let us also calculate the temperature $T(x)$ and heat flux $q^x(x)$ associated with the flow profile \eqref{exactone}. Recalling that $A=0$, and assuming that also $\chi(P,T)=\chi(P)$ is a function of the pressure alone, the first line of \eqref{dopo} becomes
\begin{equation}
\partial_x T =\dfrac{\eta \Bar{C}}{\chi} \, u \equiv -\dfrac{q^x}{\chi} \, .
\end{equation}
Using the previously found expressions for $u$ and $\Bar{C}$, and adopting the boundary conditions $T(\pm L/2)=T_+$, we obtain
\begin{equation}\label{temperatueandheattone}
\begin{split}
T={}& T_+ + \dfrac{\eta}{\chi} \ln \left[\dfrac{1}{\sqrt{1{-}v^2}} \cos\left(\dfrac{2x}{L}\arctan\left(\dfrac{v}{\sqrt{1{-}v^2}}\right) \right) \right]\, ,\\
q^x={}& \dfrac{2\eta}{L} \arctan\left(\dfrac{v}{\sqrt{1{-}v^2}}\right) \tan\left[ \dfrac{2x}{L} \arctan\left(\dfrac{v}{\sqrt{1{-}v^2}}\right)\right]\, ,
\end{split}
\end{equation}
which are plotted in Fig. \ref{fig:temperone} for various values of $v$. The physical interpretation of these profiles is rather natural: Due to viscous dissipation, the interior of the fluid tends to warm up, while its surface is kept at a fixed temperature $T_+$. As a result, in a stationary state, the system is hotter in the center, and there is a persistent flow of heat from the center to the boundaries. 

\subsection{Landau frame}
\vspace{-0.1cm}

\begin{figure}[b!]
    \centering
\includegraphics[width=0.47\linewidth]{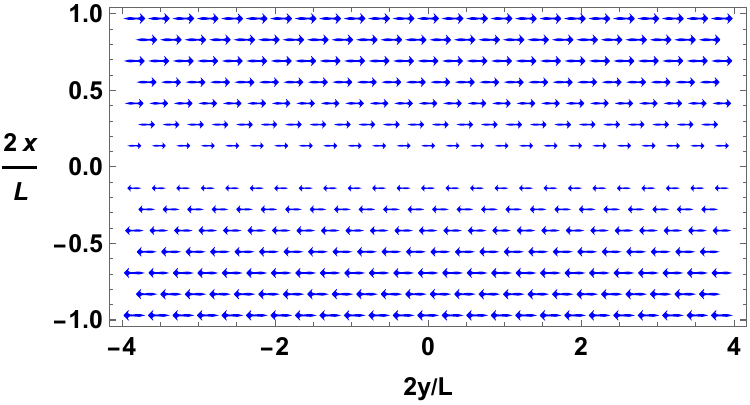}
\includegraphics[width=0.47\linewidth]{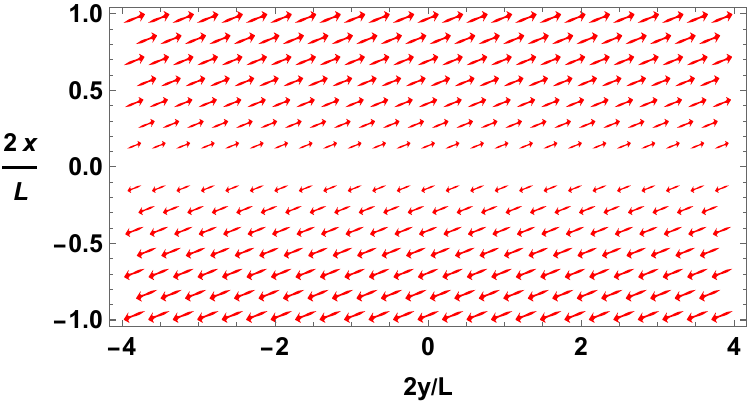}
\caption{Comparison between Eckart's flow velocity $u^j/u^t$ (left panel) and Landau's flow velocity $u_L^j/u_L^t$ (right panel) of a fluid with uniform viscosity undergoing a generalized Couette flow (see Fig.~\ref{fig:sketch}), assuming equal boundary temperatures ($T_+=T_-$) and luminal plate velocities ($v\rightarrow 1$). The analytical expressions are given by Eq.s \eqref{exactone} and \eqref{landizia}, where the ratio \eqref{theSloope} has been taken to be $0.2$, to highlight the differences (this ratio should be much smaller than $1$, for hydrodynamics to be applicable). Intuitively, the left panel describes the flow of particles, while the right panel describes the flow of energy.}
    \label{fig:Landauons}
\end{figure}

Now that we have obtained the solution in the Eckart frame, let us convert our results to the Landau frame. Generally, this conversion requires the solution of the eigenvalue problem
$T^{\mu}_{~\nu} u_{L}^{\nu} =- \epsilon_{L} u^{\mu}_{L}$ 
to determine the energy density $\epsilon_{L}$ and fluid velocity $u^{\mu}_{L}$ in the Landau frame, as the timelike eigenvalue and eigenvector \cite{landau6,OlsonLifsh1990}. However, it is well known that, to first order in the inverse Reynolds number $\text{Re}^{-1}$, which is determined by the ratio of \text{``Dissipative fluxes''} to \text{``Equilibrium fluxes''}, Landau's velocity is related to Eckart's one via \cite{Israel_Stewart_1979,GavassinoNonHydro2022}
\begin{equation}
u_L^\mu = u^\mu +\dfrac{q^\mu}{\varepsilon+P}+\mathcal{O}(\text{Re}^{-2})\, ,
\end{equation}
and we will limit our discussion to this regime.
Using Eq.s \eqref{exactone} and \eqref{temperatueandheattone}, plus the vanishing of $q^y$ and $q^z$, we find
\begin{equation}\label{landizia}
\begin{split}
u_L^t={}& \sqrt{1+(u_L^x)^2+(u_L^y)^2}\, , \\
u_L^x={}& \dfrac{2\eta}{L(\varepsilon{+}P)} \arctan\left(\dfrac{v}{\sqrt{1{-}v^2}}\right) \tan\left[ \dfrac{2x}{L} \arctan\left(\dfrac{v}{\sqrt{1{-}v^2}}\right)\right] \, , \\
u_L^y={}& \tan\left[\dfrac{2x}{L} \arctan\left(\dfrac{v}{\sqrt{1{-}v^2}} \right) \right] \, , \\
u_L^z={}& 0 \, . \\
\end{split}
\end{equation}
As foretold, the transfer of energy from the fluid to the plates causes Landau’s velocity to acquire a component perpendicular to the boundaries. Physically, this makes intuitive sense: Since the fluid is releasing heat onto the plates, and the Landau velocity tracks the energy flow, some fluid parcels should cross the interface, and be effectively ``absorbed'' by the plates (see Fig. \ref{fig:Landauons}). We can quantify the magnitude of this effect by simply taking the ratio
\begin{equation}\label{theSloope}
\dfrac{u_L^x}{u_L^y}=\dfrac{2\eta}{L(\varepsilon{+}P)} \arctan\left(\dfrac{v}{\sqrt{1{-}v^2}}\right)\, .
\end{equation}
This dimensionless number (which must be $\ll 1$ for hydrodynamics to be applicable) quantifies the ``deflection'' of the Landau streamlines relative to the Eckart ones. Let us examine how this quantity scales in various regimes. 
In an ultrarelativistic fluid, assuming $v\rightarrow 1$, Eq. \eqref{theSloope} converges to a finite value,
\begin{equation}
\dfrac{u_L^x}{u_L^y}\xrightarrow{v \to 1} \dfrac{\pi\eta}{L(\varepsilon{+}P)} \, ,
\end{equation}
and $\eta\, {\sim}\, (\varepsilon{+}P)\tau_R$ \cite{Denicol2012Boltzmann,DenicolANewWay2010xn,Denicol14Momenta2014vaa}, where $\tau_R$ is a microscopic equilibration timescale. Hence, $u_L^x/u_L^y$ equals the Knudsen number $\tau_R/L$ \cite{landau6}, up to a numerical factor of order unity. By contrast, in non-relativistic fluids, with small $v$, Eq. \eqref{theSloope} becomes
\begin{equation}
\dfrac{u_L^x}{u_L^y}\xrightarrow{|v| \ll 1} \dfrac{2\eta v}{L(\varepsilon{+}P)} \, ,
\end{equation}
and we have $\varepsilon{+}P\sim \rho$ and $\eta\sim \rho c_s^2 \tau_R$ \cite[\S 5.5]{huang_book}, where $\rho$ is the mass density and $c_s$ is the sound speed. Hence $u_L^x/u_L^y$ scales like $(\tau_R/L)\times c_s^2 v$, which is an order `` speed of light$^{-3}\,$''.

\subsection{Asymmetric case}
\vspace{-0.3cm}

We now consider the case with unequal plate temperatures, $T_+ \neq T_-$. In the nonrelativistic limit, such a temperature difference would not affect the flow profile \eqref{couetteNewtonian} (provided that $\eta(P,T)$ does not depend on $T$), since momentum conservation always gives $\eta\partial_x u = C$. However, in relativity, we know that the situation is different: The heat inertia term $u q^x$ in the third line of \eqref{ABC} couples the temperature gradient to the velocity field, making the flow profile sensitive to the difference between $T_+$ and $T_-$. Let us examine this effect in detail.

\begin{figure}[b!]
    \centering
\includegraphics[width=0.47\linewidth]{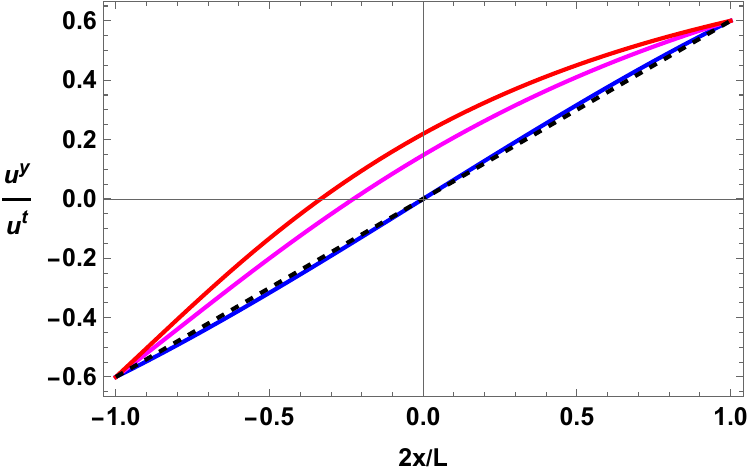}
\includegraphics[width=0.47\linewidth]{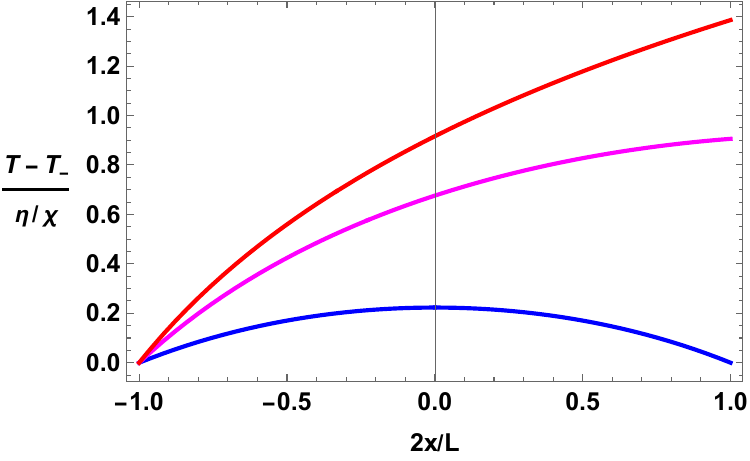}
\includegraphics[width=0.47\linewidth]{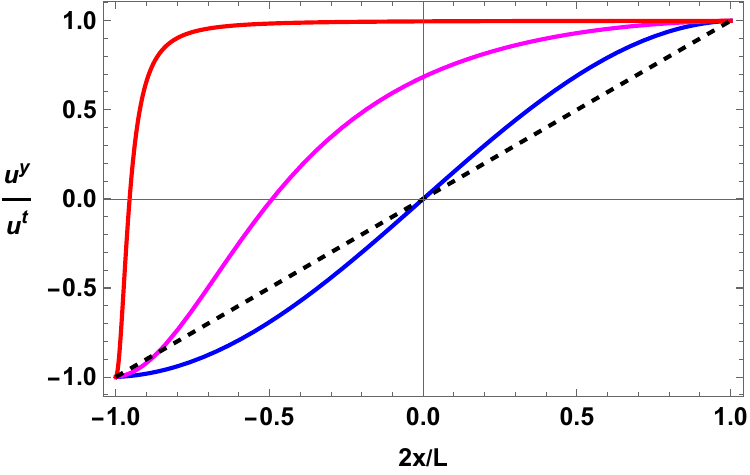}
\includegraphics[width=0.47\linewidth]{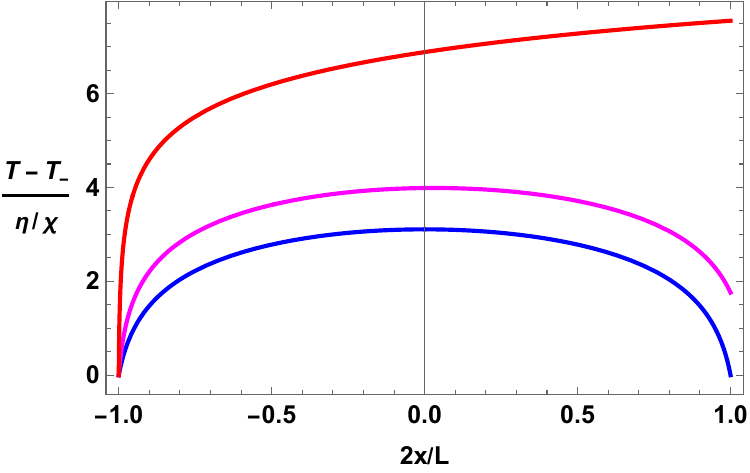}
\caption{Velocity (left column) and temperature (right column) of a relativistic fluid with uniform $\eta$ and $\chi$ undergoing a generalized Couette flow (see Fig.\ref{fig:sketch}), for unequal boundary temperatures, and with plate velocities $v=0.6$ (upper row), and $0.999$ (lower row). The analytical expressions are provided in Eq. \eqref{uasymm}, where we took $h=0$ (blue), $-1$ (magenta), $-100$ (red).}
    \label{fig:Asymmetric}
\end{figure}

When $T_+ \neq T_-$, the discrete rotational symmetry discussed in Sec.~\ref{Thelasymmetry} is broken, and $A$ can no longer be set to zero. Physically, $A \equiv T^{xt}$ represents the net energy flux from the plate at $x=-L/2$ to the plate at $x=+L/2$, which naturally increases with the temperature difference $T_- - T_+$. Hence, we need to solve \eqref{dopo} with arbitrary $A$ and $C$. Assuming $\eta$ and $\chi$ constant, we obtain
\begin{equation}\label{uasymm}
\begin{split}
u={}& h \csc\left(\dfrac{x-x_0}{l} \right) -\sqrt{1{+}h^2}\, \cot\left(\dfrac{x-x_0}{l}\right) \, , \\
T={}&T_0 -\dfrac{\eta}{\chi}\left\{\ln \left[1+\left(1-\frac{h}{\sqrt{1{+}h^2}}\right) u \left(u-\sqrt{1{+}u^2}\right)\right]+\text{arcsinh}(u)\right\} \, , \\
\end{split}
\end{equation}
where $x_0$ and $T_0$ are integration constants. Furthermore, we have introduced $l=(\Bar{C}^2{-}\Bar{A}^2)^{-1/2}$ and $h=\Bar{A}/ (\Bar{C}^2{-}\Bar{A}^2)^{1/2}$. Using the boundary condition $u(\pm L/2)=\pm v/\sqrt{1{-}v^2}$, we obtain
\begin{equation}
l= \dfrac{L/2}{\arctan\left[ \dfrac{v}{ \sqrt{(1{+}h^2)(1{-}v^2)}} \right]} \, , \qquad
x_0 = l\arccos\left[ \dfrac{h}{\sqrt{h^2 +(1{-}v^2)^{-1}}} \right]\, .
\end{equation}
In Fig.~\ref{fig:Asymmetric}, we show the solutions for velocity and temperature considering several values of $v$ and $h$. As is evident from the plots, a temperature difference between the plates breaks the odd symmetry of the velocity profile \eqref{exactone}. This asymmetry becomes more pronounced at higher $v$, since the inertia-of-heat term $u q^x$ increases with $u$.

\newpage
\section{Solution in the Landau frame}
\vspace{-0.2cm}
In the absence of a conserved charge,  there is no Eckart frame, and one must work directly in the Landau frame. For this reason, we conclude by considering the symmetric Couette configuration 
(the case with $v\,{\neq}\, 0$ and $T_+\,{=}\,T_-$) 
for an ultrarelativistic fluid without a conserved charge, with equation of state $\varepsilon=3P\propto T^{4}$. Given that the derivation is already technically demanding, we neglect the Israel–Stewart relaxation term and simply adopt  the constitutive relation $\pi^{\mu\nu}=-2\eta\Delta^{\nu\rho}_{\alpha\beta}\partial^\alpha u^\beta$. As in the previous sections, we work with constant shear viscosity, $\eta=\text{const}$. 

%This approximation is justified, as we have seen in the previous sections that Israel-Stewart corrections have only a minor influence on the overall flow structures (see, for instance, Fig.~\ref{fig:SmoothShock}).

\vspace{-0.2cm}
\subsection{Conservation laws}
\vspace{-0.2cm}

Requiring as usual that $\partial_\mu\,{=}\,(0,\partial_x,0,0)$ and  $u^\mu\,{=}\,(\gamma,u_x,u_y,0)$, the energy-momentum conservation equations give\footnote{For a fluid at zero chemical potential described in the Landau frame, the heat flux $q^\mu$ vanishes. In addition, an ultrarelativistic equation of state implies zero bulk pressure, $\Pi=0$. Therefore, the shear stress tensor $\pi^{\mu\nu}$ is the only dissipative contribution left in \eqref{TimunuF}.}
\begin{equation}\label{ABCZeroChem}
\begin{split}
\partial_x T^{xt}=0 & \qquad \Longrightarrow \qquad 4P\gamma u_x +\pi^{xt}=0 \, , \\
\partial_x T^{xx}=0 & \qquad \Longrightarrow  \qquad P(1{+}4u_x^2)  +\pi^{xx}=B \, , \\
\partial_x T^{xy}=0 & \qquad \Longrightarrow \qquad  4Pu_x u_y +\pi^{xy}=C \, .\\
\end{split}
\end{equation}
In the first line, we have directly set $A=0$ thanks to the symmetry argument of Sec. \ref{Thelasymmetry}. Using the algebraic constraint $-\pi^{xt}\gamma+\pi^{xx}u_x+\pi^{xy}u_y=0$, we can isolate $P$, $\pi^{xx}$, and $\pi^{xy}$, and we obtain
\begin{equation}
\begin{split}
\dfrac{P}{B}={}& -\dfrac{1}{3}-\dfrac{Ru_y}{3u_x} \, , \\
\dfrac{\pi^{xx}}{B}={}& \dfrac{4}{3}(1{+}u_x^2) +\dfrac{R u_y}{3u_x} (1{+}4u_x^2) \, , \\
\dfrac{\pi^{xy}}{B}={}& \dfrac{4}{3} u_x u_y +\dfrac{R}{3} (3{+}4u_y^2)  \, , \\
\end{split}
\end{equation}
where we have introduced the ratio $R=C/B$. Finally, invoking the constitutive relation $\pi^{\mu\nu} = -2\eta\,\Delta^{\nu\rho}_{\alpha\beta}\,\partial^\alpha u^\beta$, we can convert the second and third lines into coupled differential equations for $u_x$ and $u_y$, which read
\begin{equation}\label{uxuyandsoon}
\begin{split}
\partial_{\Bar{x}} u_x ={}& -1-\dfrac{Ru_y}{4u_x}\, \dfrac{1{+}4u_x^2}{1{+}u_x^2} \, , \\
\partial_{\Bar{x}} u_y ={}& -\dfrac{u_x u_y}{1{+}u_x^2}-\dfrac{R}{1{+}u_x^2} \left[1+u_y^2 +\dfrac{u_y^2}{4(1{+}u_x^2)} \right] \, , \\
\end{split}
\end{equation}
where we have introduced the dimensionless coordinate $\Bar{x}=Bx/\eta$.

\vspace{-0.2cm}
\subsection{How to solve the equations}
\vspace{-0.3cm}

\begin{figure}[h!]
    \centering
\includegraphics[width=0.46\linewidth]{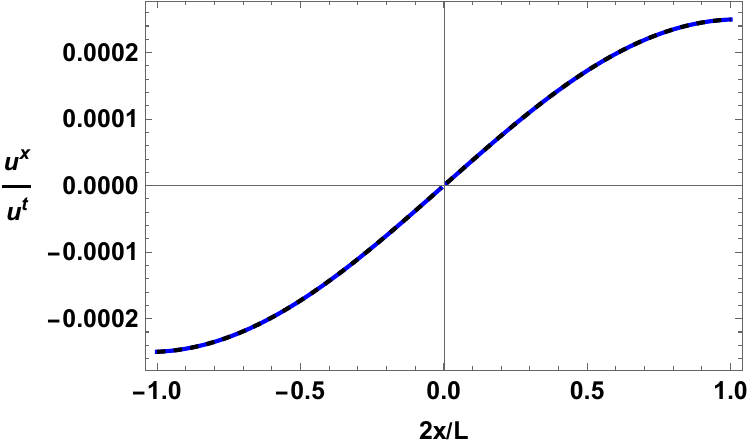}
\includegraphics[width=0.46\linewidth]{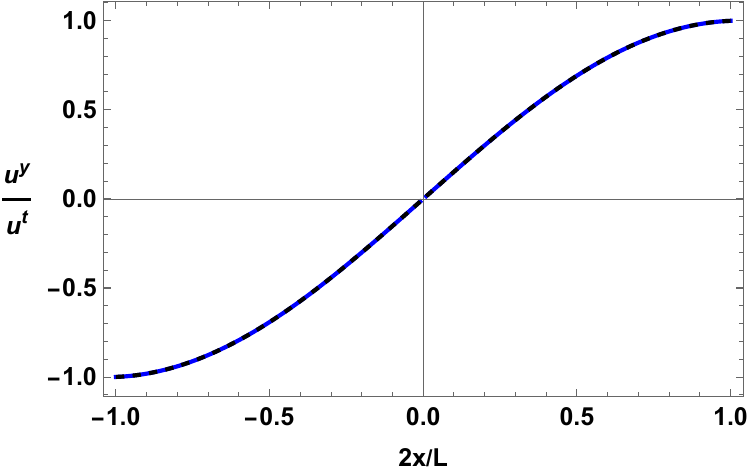}
\caption{Velocity profile of an ultrarelativistic fluid with zero chemical potential and constant $\eta$ undergoing a generalized Couette flow (see Fig. \ref{fig:sketch}) for equal boundary temperatures ($T_+\,{=}\,T_-$), $v\,{=}\,0.999$, and $R\,{=}\,{-}0.001$. The blue curve is the numerical solution of the system \eqref{uxuyandsoon}, while the dashed line is the analytical small-$R$ approximation \eqref{divergent}, which is valid as long as hydrodynamics is reliable.}
\label{fig:TheRealZerochem}
\end{figure}

Since the system is symmetric under the transformation $(x,y,z)\rightarrow ({-}x,{-}y,z)$, the flow velocity must vanish at the origin, namely $u_x(0)\,{=}\,u_y(0)\,{=}\,0$. This provides the initial data for the system \eqref{uxuyandsoon}. However, a subtlety arises, because the first line of \eqref{uxuyandsoon} becomes singular when $u_x=0$. Consequently, one must treat the limit of $\partial_{\Bar{x}}u_x$ as $\Bar{x}\rightarrow 0$ with care. In particular, as both $u_x$ and $u_y$ approach zero in this limit, we should invoke L'H\^{o}pital's rule, yielding
\vspace{-0.1cm}
\begin{equation}
\lim_{\Bar{x}\rightarrow 0} \dfrac{u_y}{u_x} \stackrel{\text{L'H}}{=} \lim_{\Bar{x}\rightarrow 0} \dfrac{\partial_{\Bar{x}} u_y}{\partial_{\Bar{x}} u_x} \, .
\end{equation}
Hence, taking the limit of \eqref{uxuyandsoon} as $\Bar{x}\rightarrow 0$, we find
\begin{equation}
\begin{split}
\partial_{\Bar{x}} u_x(0) ={}& -1-\dfrac{R \partial_{\Bar{x}} u_y(0)}{4 \partial_{\Bar{x}} u_x(0)} \, , \\
\partial_{\Bar{x}} u_y(0) ={}& -R  \, , \\
\end{split}    
\end{equation}
which can be solved to obtain
\vspace{-0.1cm}
\begin{equation}\label{polishit}
\partial_{\Bar{x}} u_x(0) = \dfrac{-1\pm \sqrt{1{+}R^2}}{2} \, .  
\end{equation}
To choose the sign, we recall that hydrodynamics is valid only under the assumption of near local equilibrium, i.e., when the pressure greatly exceeds the viscous stresses. This condition requires $|R|\ll 1$ and $|\partial_{\Bar{x}} u_x|\ll 1$, which dictates the choice of the positive sign in \eqref{polishit}. Expanding for small $R$ and $\Bar{x}$, we arrive at the following approximation:
\vspace{-0.1cm}
\begin{equation}\label{uxuysolutionapprox}
\begin{cases}
u_x \approx  R^2\Bar{x}/4 \, , \\
u_y \approx   -R\Bar{x} \, . \\
\end{cases} \spc (\text{small }R\text{ and }\Bar{x}) \, .
\end{equation}
We can, however, do better than this. In the limit of small $R$, we have that $u_x^2 \ll 1$ at all $\Bar{x}$. Hence, the first line of \eqref{uxuyandsoon} simplifies to $u_x=-Ru_y/4$, and the second line simplifies to $\partial_{\Bar{x}} u_y =-R(1+u_y^2)$. The resulting solution is
\vspace{-0.1cm}
\begin{equation}\label{divergent}
\begin{cases}
u_x \approx  R \tan(R\Bar{x})/4 \, , \\
u_y \approx   -\tan(R\Bar{x}) \, . \\
\end{cases} \spc (\text{small }R) \, .
\end{equation}
If $R$ is sufficiently small, the above approximation remains accurate for all values of $\Bar{x}$, up to $|\Bar{x}| = |R^{-1}|\pi/2$, where the fluid velocity formally reaches the speed of light. In practice, however, the motion of the plates is always subluminal. Consequently, the system \eqref{divergent} must be supplemented with the boundary conditions $u_y(x{=}\pm L/2)=\pm v/\sqrt{1{-}v^2}$, which fix the value of $R$ and define the physical range of $\Bar{x}$:
\vspace{-0.1cm}
\begin{equation}\label{RandR}
\begin{split}
& R\approx -\dfrac{2\eta}{LB} \arctan\left(\dfrac{v}{\sqrt{1{-}v^2}} \right)\, , \\ 
& \Bar{x}_{\text{plate }\pm} \equiv  \pm \dfrac{LB}{2\eta} \approx \pm   |R^{-1}| \arctan\left(\dfrac{v}{\sqrt{1{-}v^2}} \right) \, .\\
\end{split}
\end{equation}
In Fig. \ref{fig:TheRealZerochem}, we provide a quantitative example, where we solve \eqref{uxuyandsoon} numerically and compare the result with the approximation \eqref{divergent} (which is clearly very accurate over the entire domain). Note that, within the approximation \eqref{divergent}, one has $P\equiv B$. Indeed, for the configuration shown in figure \ref{fig:TheRealZerochem}, the residual difference between $P$ and $B$ is found to be comparable to the numerical discretization error.

\vspace{-0.3cm}
\subsection{Including a conserved charge}\label{lastissimo}
\vspace{-0.2cm}

We conclude with an interesting observation. Consider a non-degenerate ultrarelativistic gas, with a conserved number density $n$, whose equation of state is given by $\varepsilon\,{=}\,3P\,{=}\,3nT$ \cite[\S 2.4.4]{rezzolla_book}. In the Landau frame, the stress–energy tensor retains the form $T^{\mu \nu}=4Pu^\mu u^\nu +Pg^{\mu \nu}+\pi^{\mu \nu}$ (with no bulk term $\Pi$, see \cite{Weinberg1971,BulkGavassino}), implying that \eqref{ABCZeroChem} remains valid also for this system. Consequently, the equations \eqref{uxuyandsoon} also govern the Landau-frame Couette flow of such a fluid.

At the same time, because this fluid possesses a conserved charge, the Eckart frame exists, and the analysis presented in Sec.~\ref{finitUdine} applies as well. This means that one may also determine the flow profile in the Eckart frame, and subsequently transform it to the Landau frame. The resulting solution is the flow \eqref{landizia}, which happens to coincide with the approximation \eqref{divergent}-\eqref{RandR}.
Therefore, Fig.~\ref{fig:TheRealZerochem} may also be interpreted as a comparison between a flow profile obtained directly in the Landau frame (blue~line) and that obtained by solving in the Eckart frame and later transforming to the Landau frame (dashed~line). As expected, in the hydrodynamic regime (i.e. $|R|\ll 1$ and $\partial_{\Bar{x}}u_x \ll 1$), the two results coincide, confirming the physical equivalence of the two hydrodynamic frames at small gradients \cite{Israel_Stewart_1979,Kovtun2019,GavassinoNonHydro2022,GavassinoUniveraalityI2023odx}.

\section{Conclusions}

In this work, we have constructed and analyzed a class of stationary, planar-symmetric solutions of relativistic hydrodynamics representing Couette-type flows between parallel plates. Our results clarify the fundamental role of the heat flux in determining the structure of such configurations. In the relativistic regime, the heat flux contributes directly to the momentum density, and cannot be consistently neglected even when the viscosity is uniform. The omission of this term, as in earlier treatments \cite{Rogava1996}, leads to qualitatively incorrect flow profiles that overestimate the velocity shear and violate energy conservation (see Fig. \ref{fig:speeddoni}).

For fluids at finite chemical potential modeled in the Eckart frame, the generalized Couette problem admits exact analytic solutions that remain regular in the ultrarelativistic limit, and reduce smoothly to the Newtonian form at low velocities. The resulting temperature and heat-flux profiles tell a rather intuitive physical story: viscous dissipation heats the interior, while the generated entropy is removed by a steady flow of energy toward the plates (see Fig.~ \ref{fig:temperone}). For this reason, when the analysis is transformed from the Eckart to the Landau frame, one finds that the fluid parcels, which now follow the flow of energy, tend to cross the boundaries and be ``absorbed'' by the plates (see Fig.~ \ref{fig:Landauons}). This phenomenon notably complicates the analysis in fluids without a conserved charge, since for these fluids there is no Eckart frame, and one needs to work directly in the Landau frame. Nevertheless, the final flow profile is essentially the same as in fluids with a conserved charge (see Fig.~ \ref{fig:TheRealZerochem}, and discussion is Sec. \ref{lastissimo}). We expect analogous mechanisms to operate in other stationary shear-flow configurations, such as Poiseuille flow.

%At zero chemical potential, qualitatively new phenomena emerge when the plates are at rest but maintained at different temperatures. In this regime, the pressure depends exclusively on the temperature, so any thermal gradient necessarily induces motion. As a consequence, no stationary hydrodynamic profile can smoothly connect two distinct boundary temperatures without developing shocks. Therefore, the only stationary configuration consistent with energy-momentum conservation is one in which the fluid remains uniform between the plates. This, in turn, implies that the temperature of the fluid cannot match that of the plates at the interfaces. The temperature adjustment must instead occur within thin, non-equilibrium layers inside the solid boundaries, where the fluid particles experience their first interactions with the material constituents of the plates. These transition layers are analogous to stellar photospheres, and can be accurately modeled within relativistic kinetic theory (see Fig.~\ref{fig:TheRightTxtTxxNoChem}).

All in all, our results showcase the interplay between (a) the inertia of heat, (b) viscous heating, and (c) boundary physics. In a relativistic setting, the simultaneous action of these ingredients generates behavior with no classical analogue, providing a controlled arena in which to probe the thermodynamic organization of relativistic shear flows. In particular, we uncover a general mechanism whereby the energy produced by viscous dissipation is compensated by transport toward the boundaries. In the Eckart frame, this balance is naturally expressed in terms of a transverse heat flux (which modifies the longitudinal speed via inertia of heat), while in the Landau frame the same physics manifests itself through a transverse perturbation of the velocity field, without an explicit heat current. Although the two descriptions are equivalent within the accuracy of the gradient expansion, the choice of hydrodynamic frame governs how this compensation mechanism is represented and how physically meaningful boundary conditions are implemented in steady states. From this perspective, frame transformations do not merely reshuffle variables, but also reshape physical interpretation. Let us stress, however, that realistic Couette configurations are expected to become turbulent at sufficiently large values of \(v\), and the stationary solutions constructed here should therefore be regarded as idealized benchmarks.

We finally remark that, throughout this work, we have implicitly assumed that the fluid temperature and tangential velocity match those of the plates, i.e. we have imposed the usual no-slip boundary conditions.
However, one important situation in which this assumption ceases to be valid is the case of a fluid without conserved charges, confined between parallel plates held at different temperatures, $T_+ \neq T_-$. Since this configuration, not addressed here, requires a more refined treatment, we will return to this problem in a forthcoming paper.

% There is, however, one important situation in which this assumption ceases to be valid: a zero–chemical–potential fluid confined between plates held at different temperatures, $T_+ \neq T_-$. This configuration, not addressed here, requires a more refined treatment. We will return to this problem in a forthcoming paper, where we show that its consistent formulation necessarily involves the use of kinetic theory near the interface.

\section*{Acknowledgements}

This work was supported in part by the Deutsche Forschungsgemeinschaft (DFG, German Research Foundation) through the CRC-TR 211 ‘Strong-interaction matter under extreme conditions’-project number 315477589 – TRR 211, and DFG research grant TR 1733/3-1 ‘Open quantum systems in Euclidean and real-time approaches’ - project number 523702416. LG is supported by a MERAC Foundation prize grant, an Isaac Newton Trust Grant, and funding from the Cambridge Centre for Theoretical Cosmology. G.~S.~D.~acknowledges funding from CNPq.

\appendix

\section{Radiative transport coefficients}\label{AppAAA}

For completeness, let us consider at least one example where the transport coefficients depend on the temperature. A particularly convenient case is that of a mixed matter-radiation system, where the matter behaves as an ideal fluid and the radiation quanta have a finite, constant mean free path $\tau$. Under the grey-opacity approximation, the transport coefficients of the combined matter–radiation fluid are then given by \cite{Weinberg1971,UdeyIsrael1982,GavassinoRadHydroDisp2024cqw}
\begin{equation}
\begin{split}
\eta={}& \dfrac{4}{15} aT^4 \tau \, , \\
\chi={}& \dfrac{4}{3} aT^3 \tau \, . \\
\end{split}
\end{equation}
This fluid has the notable property that $d\eta/dT=4\chi/5$. Hence, the first line of \eqref{dopo} can be rewritten as a differential equation for $\eta$, rather than $T$. Under the geometric assumptions of Sec. \ref{Thelasymmetry}, we therefore obtain
\begin{equation}\label{dopoRAD}
\begin{split}
\partial_x \eta={}& \dfrac{4}{5} Cu \, ,\\
\partial_x u={}& -\dfrac{C(1{+}u^2)}{\eta} \, .\\
\end{split}
\end{equation}
The trick here is to take the ratio between the two equations, and to interpret the result as a separable ordinary differential equation for the function $\eta(u)$. The solution is
\begin{equation}
\eta=\dfrac{D}{(1{+}u^2)^{2/5}} \, ,
\end{equation}
with $D$ an integration constant. Plugging this formula into the second line of \eqref{dopoRAD}, we obtain a separable differential equation for the function $u(x)$, whose solution is, accounting for the boundary conditions,
\begin{equation}
f(u)=\dfrac{2x}{L} f\left(\dfrac{v}{\sqrt{1{-}v^2}} \right)\, .
\end{equation}
Here, we have introduced the function
\begin{equation}
f(u)\equiv \int_0^u \dfrac{d\xi}{(1+\xi^2)^{7/5}}=u \, _2 F_1 \left(\dfrac{1}{2},\dfrac{7}{5};\dfrac{3}{2};-u^2 \right) \approx -\dfrac{\sqrt{\pi}\, \Gamma(-1/10)}{8\,\Gamma(2/5)} \dfrac{u}{\sqrt{1+u^2}}\, ,
\end{equation}
where $_2 F_1$ is the hypergeometric function. The approximation given in the final expression is based on a numerical fit, which remains accurate to within about 3\% over the entire range of $u$, and becomes exact in the limit $u \to \infty$. Thanks to this approximation, the solution acquires a remarkably simple form:
\begin{equation}
\begin{split}
u\approx{}& \dfrac{2vx/L}{\sqrt{1-(2vx/L)^2}}\, ,\\
T\approx{}& T_+ \left[ \dfrac{1-(2vx/L)^2}{1-v^2}\right]^{1/10}\, .\\
\end{split}
\end{equation}

\bibliography{Biblio}

\label{lastpage}
\end{document}